\documentclass[english,aps,prl,notitlepage,twocolumn]{revtex4-1}
\usepackage[T1]{fontenc}
\usepackage[latin9]{inputenc}
\usepackage{color}
\usepackage{amstext}
\usepackage{amssymb}
\usepackage{graphicx}
\usepackage{esint}

\makeatletter

\usepackage{amsmath}
\newcommand\ket[1]{\left| #1\right\rangle}
\newcommand\bra[1]{\left\langle #1\right|}
\newcommand\braket[2]{\left\langle #1| #2\right\rangle }

\makeatother

\usepackage{babel}
\begin{document}
\global\long\def\braket#1#2{\left\langle #1| #2\right\rangle }

\title{Symmetry protected single photon subradiance}

\author{Han Cai$^1$, Da-Wei Wang$^1$, Anatoly A. Svidzinsky$^1$, Shi-Yao Zhu$%
^2$ and Marlan O. Scully$^{1,3,4}$}

\affiliation{$^1$Texas A\&M University, College Station TX 77843; $^2$Beijing
Computational Science Research Center, Beijing, China; $^3$Princeton
University, Princeton NJ 08544; $^4$Baylor University, Waco, TX 76706}
\begin{abstract}
We study the protection of subradiant states by the symmetry of the
atomic distributions in the Dicke limit, in which collective Lamb
shifts cannot be neglected. We find that anti-symmetric states are
subradiant states for distributions with reflection symmetry. Continuous
symmetry can also be used to achieve subradiance. This study is relevant
to the problem of robust quantum memory with long storage time and
fast readout.
\end{abstract}

\pacs{\textcolor{black}{42.50.Nn}}

\maketitle
\textcolor{black}{Cooperative spontaneous emission (Dicke superradiance
\cite{DickePR1954}) and the cooperative vacuum induced levels shifts
(Lamb shifts \cite{JLWEPR1947}) are hot topics in quantum optics.
For extended ensembles when the size of the atomic cloud is much larger
than the wavelength, the directional emission \cite{ScullyPRL2006,CrubellierOC1980}
and collective Lamb shift \cite{RohlsbergerFderP2013} of single photon
superradiance \cite{VanderWalScience2003,BlackPRL2005,KuzmichNature2003,BalicPRL2005,ScullyPRL2006,AnatolyPRA2008,AnatolyPRA2010,WangPRL2015,Svid15}
have attracted much interest. Recently it has been shown \cite{ScullyPRL2015}
that it is possible to use subradiance (the cooperative suppression
of spontaneous emission \cite{Svid15}) to store a photon in a small
volume for many atomic lifetimes; and later switch the subradiant
state to a superradiant state which emits a photon in a small fraction
of an atomic lifetime. Such a process has potential applications in
e.g., quantum informatics.}

\textcolor{black}{It has been proved that the distribution of the
atoms (e.g., periodic or random) in an extended ensemble has a substantial
effect on cooperative spontaneous emission \cite{FengPRA2014}. However,
the effect of the atomic distribution in the Dicke limit has been
studied only a little. Since the distance between atoms is much smaller
than the wavelength, one might guess that the distribution of atoms
is not important. We here show that the collective Lamb shift cannot
be neglected in general. However, by analyzing the relation between
the symmetry of the atomic distribution and cooperative emission,
we demonstrate the mitigation of the collective Lamb shift and the
symmetry protected subradiance. }

\textcolor{black}{The $N$-atom sample (size much smaller compared
to the transition wavelength $\lambda$) excited by a single photon
can be described by the Dicke state
\begin{equation}
\left|+\right\rangle =\frac{1}{\sqrt{N}}\sum\limits _{j=1}^{N}\left|j\right\rangle ,\label{eq:+state}
\end{equation}
where $\left|j\right\rangle =\left|b_{1},b_{2}...a_{j}...b_{N}\right\rangle $,
$a_{j}$($b_{j}$) is the excited (ground) state of the $j$th atom.
Since the size of atom is much smaller than the wavelength of the
coupling field, we could use dipole approximation. It also allows
us to distribute many atoms within one wavelength, which is called
the Dicke limit. The probability amplitude of the state (\ref{eq:+state})
decays at the rate $\Gamma_{+}=N\gamma$ where $2\gamma$ is the single
atom population decay rate. In the ``opposite'' case, if we neglect
Lamb shift, the single photon subradiance state
\begin{equation}
\left|-\right\rangle =\frac{1}{\sqrt{N}}(\sum\limits _{j=1}^{N\text{/2}}\left|j\right\rangle -\sum\limits _{j=N/2+1}^{N}\left|j\right\rangle ),\label{eq:-state}
\end{equation}
does not decay, i.e., $\Gamma_{-}=0$ because of the destructive interference
of the atomic transitions. However, when the cooperative Lamb shifts,
i.e., the effects of emission and reabsorption of virtual photons,
are counted in, it can degrade superradiance \cite{FriedbergPhysRepo1973,RohlsbergerFderP2013,ScullyScience2010}.
In single photon superradiance, this does not overwhelm the collective
enhancement of spontaneous emission. Not so in the case of subradiance,
the collective Lamb shift can now destroy the ability of the atoms
to ``store'' light, i.e., the original subradiant states are not
necessarily subradiant anymore. For random atomic distribution, since
each atom ``sees'' different neighboring atoms, collective Lamb
shift type fluctuation induced dephasing significantly degrades the
destructive interference.}

\begin{figure}
\textcolor{black}{\includegraphics[width=1\columnwidth]{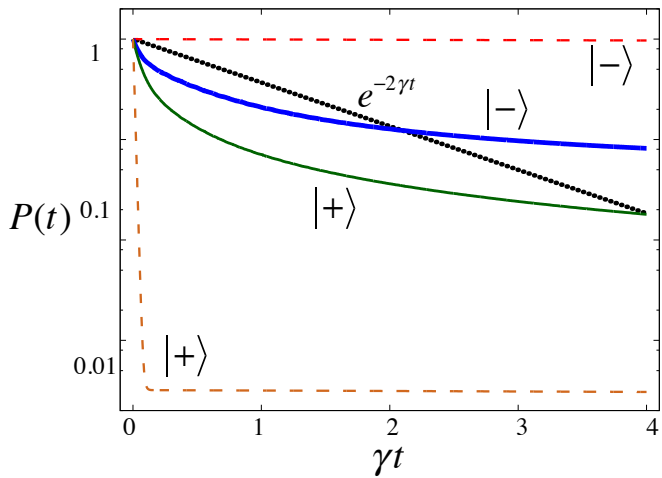}\caption{Probability $P(t)=\protect\braket{\Psi(t)}{\Psi(t)}$ to find atoms
excited as a function of time for atoms initially prepared in the
$\ket+$ and $\ket-$ states. The solid curve takes the cooperative
Lamb shift into consideration, causing rapid decay for both $\ket+$
and $\ket-$. The dashed curve ignores the Lamb shift. For comparison,
we also plot single-atom decay curve $e^{-2\gamma t}$ (dot line).}
\label{fig:collectiveLambshift}}
\end{figure}

\textcolor{black}{We first turn to a more detailed study of the lifetime
of the $\ket-$ state and the way in which collective Lamb shift type
fluctuations influence the state evolution. Numerically calculated
population decay of the anti-symmetric $\left|-\right\rangle $ state
with and without taking into account virtual transitions is compared
in Fig.\ref{fig:collectiveLambshift}. The Dicke limit ensemble of
100 atoms are randomly distributed along a 1D line within $0.01\lambda$,
where $\lambda$ is the atomic transition wavelength. Fig.\ref{fig:collectiveLambshift}
shows that collective Lamb shifts $\Omega_{ij}$ degrade subradiance
of the state $\ket-$. Without Lamb shifts, $\ket-$ is subradiant.
Counting in the Lamb shifts, $\ket-$ is composed by both superradiant
and subradiant eigenstates. The components of the superradiant eigenstate
decay fast, leaving slowly decaying subradiant components.}

\textcolor{black}{The simulation is performed in the basis of single-photon
eigenstate. The eigenstates decay exponentially, i.e., $\ket{\psi_{n}(t)}=\sum_{j}\beta_{j}e^{-\Lambda_{n}t/\hbar}\ket j$,
with $\Lambda_{n}$ being the $n$th complex eigenvalue and $\beta_{j}$
being the probability amplitude to find atom $j$ excited. The eigenvalue
equations are\cite{AnatolyPRA2010,BienaimFdP2013}
\begin{equation}
\Lambda_{n}\beta_{i}=\gamma\beta_{i}-i\sum\limits _{i\neq j}^{N}(-\Omega_{ij}+i\gamma_{ij})\beta_{j}=\sum\limits _{j}^{N}M_{ij}\beta_{j},\label{eq:matrix}
\end{equation}
where $\Omega_{ij}=-\frac{\cos(k_{0}r_{ij})}{k_{0}r_{ij}}\gamma$,
with $k_{0}$ the transition wave vector and $r_{ij}$ the distance
between atoms $i$ and $j$, is the collective Lamb shift, $\gamma_{ij}=\frac{\sin(k_{0}r_{ij})}{k_{0}r_{ij}}\gamma$
is the collective decay rate, and $M_{ij}=\gamma\delta_{ij}+(i\Omega_{ij}+\gamma_{ij})(1-\delta_{ij})$
with $\delta_{ij}$ the Kronecker delta function are the elements
of the evolution matrix $\mathbf{M}$. This result is based on a scalar
radiation field, which can be regarded as the average effect of the
vector field \cite{Agarwalbook1974,akkermansPRL2008}. We use scalar
field throughout this paper for simplicity. Results for vector field
are shown in Appendix A. There is no essential difference between
the results of scalar and vector field.}
\textcolor{black}{Numerically calculating all eigenvalues $\Lambda_{n}$
by diagonalization of the matrix $\mathbf{\mathbf{M}}$, we obtain
$\ket{\Psi(t)}=\sum_{n}c_{n}e^{-\Lambda_{n}t}\ket{\psi_{n}}$, here
$c_{n}=\braket{\psi_{n}^{T}}{\Psi(0)}$ is the projection of initial
state to the single-photon Dicke-Lamb eigenstate, and $\bra{\psi_{n}^{T}}$
is the transpose of $\ket{\psi_{n}}$ (since the matrix $\mathbf{M}$
is symmetric instead of Hermitian).}

\textcolor{black}{The mitigation of the collective Lamb shifts by
arranging the atom distribution in a ring has been found useful in
maintaining superradiance \cite{GrossPhysRepo1982}. If the atoms
are distributed randomly, the transition frequencies of atoms are
different due to the different environment of each atom, superradiance
is destroyed. However, if they are arranged periodically on a ring,
all atoms have the same environment and the superradiance is recovered.
It sheds light on the importance of the symmetry of the atomic distribution. }

\textcolor{black}{Symmetry has long been investigated as a central
feature of superradiance \cite{CrubellierOC1980}. In Dicke's original
paper \cite{DickePR1954}, it was noted that the most decaying excited
state of the collective atomic distribution must be symmetric since
the ground state is symmetric and the Hamiltonian preserves symmetry.
The symmetry of the atomic distribution determines the symmetry of
the eigenstates. For the sake of simplicity, we take 1D atomic distribution
preserving reflection symmetry for example. We set $z$ along the
line of atoms and $z=0$ as the middle point of the atomic ensemble.
The mirror reflection operator $\pi$, which transforms $z\rightarrow-z$,
commutes with the matrix $\mathbf{M}$ for a periodic distribution
of atoms, $[\mathbf{M},\pi]=0$. A nondegenerate eigenstate of $\mathbf{M}$
is also an eigenstate of $\pi$ \cite{sakurai2014modern}. $N$ eigenstates
of $N$ atoms excited by a single photon are separated into two groups
with opposite eigenvalues of $\pi$, i.e., $N/2$ symmetric and $N/2$
anti-symmetric states.}

\textcolor{black}{In the Dicke limit, the ensemble size is much smaller
than the transition wavelength. If we neglect the collective Lamb
shifts $\Omega_{ij}$ in Eq.(\ref{eq:matrix}), we obtain $M_{ij}=\gamma$
for all $i$ and $j$. In this case the eigenvalues of $\mathbf{\mathbf{M}}$
are $\Lambda_{1}=N\gamma$ and all others are equal to zero, i.e.,
there is one superradiant state and $N-1$ subradiant states \cite{akkermansPRL2008}.
The superradiant eigenstate is the symmetric state $\ket+$. Any state
orthogonal to this state is subradiant, for example, the anti-symmetric
state $\ket-$. With the presence of $\Omega_{ij}$, $\ket-$ is not
subradiant any more, as shown in Fig.\ref{fig:collectiveLambshift}. }

\textcolor{black}{We can recover the subradiant nature of $\ket-$
state by rearranging atoms such that their distribution possesses
reflection symmetry, i.e., $z_{j}=-z_{N+1-j}$ and $\pi^{\dagger}\mathbf{M}\pi=\mathbf{M}$.}
\begin{figure}
\textcolor{black}{\includegraphics[width=1\columnwidth]{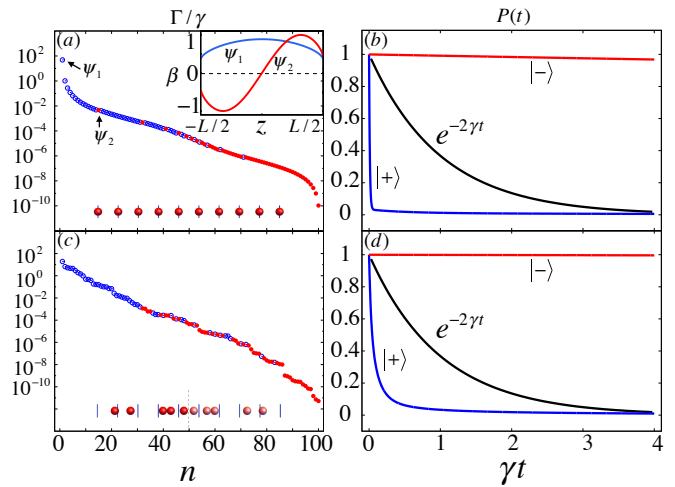}\caption{(a) Distribution of decay rates for eigenstates of an ensemble of
$100$ atoms regularly placed along a line with spacing between adjacent
atoms $0.0001\lambda$. Blue empty dots are symmetric states while
red solid dots are anti-symmetric states. Inset: Probability amplitude
$\beta_{j}$ as a function of the coordinate of the $j$th atom $z_{j}$
for the fastest decaying symmetric (blue line) and anti-symmetric
(red line) states. (b) Population decay of states $\ket+$ and $\ket-$
as a function of time. Single-atom exponentially decaying curve is
shown for comparison. (c) and (d): The same as in (a) and (b), but
for random spatial distribution of atoms with reflection symmetry.}
\label{fig:symmetryanddecayrate}}
\end{figure}
\textcolor{black}{In Fig.\ref{fig:symmetryanddecayrate}(b) we plot
the population decay for periodic distribution of atoms. The decay
of $\ket+$ is enhanced compared with the case of random distribution,
which is consistent with Ref. \cite{GrossPhysRepo1982}. On the other
hand, the decay of $\ket-$ state is drastically inhibited.}

\textcolor{black}{To analyze the reason of this inhibition, we plot
$\beta_{j}$ for the superradiant state $\ket{\psi_{1}}$ and for
a subradiant state $\ket{\psi_{2}}$ in the inset of Fig.\ref{fig:symmetryanddecayrate}(a).
It is clear that the state $\ket{\psi_{1}}$ is symmetric with respect
to the center of the sample. There is only one superradiant state
$\ket{\psi_{1}}$ with decay rate $\sim N\gamma$, as shown in Fig.\ref{fig:symmetryanddecayrate}(a).
The anti-symmetric state $\ket-$ is orthogonal to the superradiant
state $\ket{\psi_{1}}$. Because of the Dicke limit, the superradiant
state $\ket{\psi_{1}}$ shown in the inset of Fig.\ref{fig:symmetryanddecayrate}(a)
is similar to a uniform probability amplitude state $\ket+$.}

\textcolor{black}{Since reflection symmetry is the key point in the
above analysis, it is not necessary for atoms to be periodically distributed
to make the $\ket-$ state subradiant. In Fig.\ref{fig:symmetryanddecayrate}(c),
we allow half of the atoms to be distributed randomly, but in reflection
symmetry with the other half. The population decay of $\ket-$ state
is still substantially inhibited as shown in Fig.\ref{fig:symmetryanddecayrate}(d).
This is because the superradiant state $\ket{\psi_{1}}$ for random
atomic distribution is still symmetric and has no overlap with the
$\ket-$ state.}

\textcolor{black}{Generally, in the Dicke limit, we have one superradiant
and $N-1$ subradiant eigenstates. Atomic distribution determines
the symmetry of the superradiant state. By preparing atoms in an orthogonal
state to this superradiant eigenstate, we can reach subradiance and
store the photon. To release the photon, we can coherently change
the state to have the same symmetry as the superradiant eigenstate
and achieve a rapid readout \cite{ScullyPRL2015}. }

\textcolor{black}{We could achieve subradiance in extended sample
as well. For an extended spherical sample, we find \cite{ScullyPRL2015}
approximately decay rates $\Gamma_{+}^{\vec{k}_{0}}\cong\gamma[1+\frac{3}{8\pi}\frac{\lambda^{2}}{A}(N-1)]$
and $\Gamma_{-}^{\vec{k}_{0}}\cong\gamma[1-\frac{3}{8\pi}\frac{\lambda^{2}}{A}]$
for states $\ket{\pm}_{\vec{k}_{0}}=\sum_{j=1}^{N/2}e^{i\vec{k}_{0}\cdot\vec{r}_{j}}\ket{j}\pm\sum_{j=N/2+1}^{N}e^{i\vec{k}_{0}\cdot\vec{r}_{j}}\ket{j}$,
where $\lambda$ is the transition wavelength, $R$ is the radius
of the atomic cloud and $A=\pi R^{2}$ is the cross section area.
The ``extra'' $\gamma$ in $\Gamma_{+}^{\vec{k}_{0}}$ is not important,
as it is small compared to the leading term going as $\frac{3}{8\pi}\frac{\lambda^{2}}{A}N$.
However the $\gamma$ term in $\Gamma_{-}^{\vec{k}_{0}}$ is important.
It seems like for the $\left|-\right\rangle _{\vec{k}_{0}}$ state,
the single atom spontaneous decay rate is a lower decay limit for
an extended sample. The good news however is that the collective spontaneous
decay can also be mitigated by the spatial symmetry of the atomic
distribution. }In order to calculate the the evolution of atomic system
of a dense cloud of volume $V$ , we use equation with exponential
kernel \cite{AnatolyPRA2008}
\begin{equation}
\frac{\partial\beta(t,\mathbf{r})}{\partial t}=i\gamma\int d\mathbf{r'}n(\mathbf{r'})\frac{\exp(ik_{0}\left|\mathbf{r}-\mathbf{r'}\right|)}{k_{0}\left|\mathbf{r}-\mathbf{r'}\right|}\beta(t,\mathbf{r'}),\label{eq:kernal}
\end{equation}
where $\beta(t,\mathbf{r})$ is the probability amplitude to find
atom at position $\mathbf{r}$ excited at time $t$, $n(\mathbf{r})$
is the atomic density. Eq.(\ref{eq:kernal}) is valid in Markovian
(local) approximation and is the continuous limit of Eq.(\ref{eq:matrix}).
Eigenfunctions of Eq.(\ref{eq:kernal}) are $\beta(t,\mathbf{r})=e^{-\Lambda t}\beta(\mathbf{r})$
and the eigenvalues $\Lambda$ determine the evolution of the atomic
system. $\text{Re}(\Lambda)$ yields the state decay rate, while $\text{Im(}\Lambda)$
describes frequency (Lamb) shift of the collective excitation. The
eigenfunction equation for $\beta(\mathbf{r})$ reads
\begin{equation}
-i\gamma\int d\mathbf{r'}n(\mathbf{r'})\frac{\exp(ik_{0}\left|\mathbf{r}-\mathbf{r'}\right|)}{k_{0}\left|\mathbf{r}-\mathbf{r'}\right|}\beta(\mathbf{r'})=\Lambda\beta(\mathbf{r}).\label{eq:eigenfunction in appendix}
\end{equation}
We consider an infinitely long cylindrical shell of radius $R$ and
use cylindrical coordinates $\mathbf{r}=(\rho,\varphi,z)$. The atomic
density is $n(\mathbf{r})=n_{0}\delta(\rho-R)/2\pi R$, where $n_{0}$
is the number of atoms per unit length of the cylinder. For such geometry
Eq.(\ref{eq:eigenfunction in appendix}) reads 
\begin{equation}
-\frac{i\gamma n_{0}}{2\pi}\int\limits _{0}^{2\pi}d\varphi'\int\limits _{-\infty}^{\infty}dz'K(\varphi-\varphi',z-z')\beta(\varphi',z')=\Lambda\beta(\varphi,z),\label{eq:eqA4}
\end{equation}
where
\[
K(\varphi,z)=\frac{\exp[ik_{0}\sqrt{2R^{2}-2R^{2}cos\varphi+z^{2}}]}{k_{0}\sqrt{2R^{2}-2R^{2}cos\varphi+z^{2}}}.
\]
We look for solution of Eq.(\ref{eq:eqA4}) in the form 
\begin{equation}
\beta(\varphi,z)=e^{in\varphi}e^{ik_{z}z},\label{eq:beta}
\end{equation}
where $n$ is an integer number and $k_{z}$ is the wave number of
the mode along the cylindrical axis $z$. Substituting Eq.(\ref{eq:beta})
in Eq.(\ref{eq:eqA4}) we obtain the following equation for eigenvalues
$\Lambda_{n}$
\begin{equation}
\Lambda_{n}=-\frac{i\gamma n_{0}}{2\pi}\int\limits _{0}^{2\pi}d\varphi'\int\limits _{-\infty}^{\infty}dz'K(\varphi',z')e^{in\varphi'}e^{ik_{z}z'}.\label{eq:eqA6}
\end{equation}
Integrating over $z'$ can be done by using the integral
\begin{equation}
\int\limits _{-\infty}^{\infty}dz'\frac{exp[ik_{0}\sqrt{r^{2}+z'^{2}}]}{\sqrt{r^{2}+z'^{2}}}e^{ik_{z}z'}=i\pi H_{0}^{(1)}(r\sqrt{k_{0}^{2}-k_{z}^{2}}),
\end{equation}
where $H_{0}^{(1)}(x)$ is the Hankel function. Then Eq.(\ref{eq:eqA6})
reduces to 
\begin{equation}
\Lambda_{n}=\frac{\gamma n_{0}}{2k_{0}}\int\limits _{0}^{2\pi}d\varphi'H_{0}^{(1)}(R\sqrt{2-2\cos\varphi'}\sqrt{k_{0}^{2}-k_{z}^{2}})e^{in\varphi'}.\label{eq:eqA8}
\end{equation}
The integration over $\varphi'$ can be calculated using
\begin{equation}
\int\limits _{0}^{2\pi}d\varphi'H_{0}^{(1)}(a\sqrt{2-2\cos\varphi'})e^{in\varphi'}=2\pi J_{n}(a)H_{n}^{(1)}(a),
\end{equation}
and Eq.(\ref{eq:eqA8}) leads to
\begin{equation}
\Lambda_{n}=\frac{\pi\gamma n_{0}}{k_{0}}J_{n}(R\sqrt{k_{0}^{2}-k_{z}^{2}})H_{n}^{(1)}(R\sqrt{k_{0}^{2}-k_{z}^{2}}).
\end{equation}

Hankel functions can be written as a combination of the Bessel functions
of the first and the second kind as
\begin{equation}
H_{n}^{(1)}(x)=J_{n}(x)+iY_{n}(x),
\end{equation}
which yields the following answer for the real and imaginary parts
of the eigenvalues $\Lambda_{n}$ for $k_{z}\leq k_{0}$\textcolor{black}{
\begin{equation}
\Gamma_{n}=\text{Re}(\Lambda_{n})=\frac{\pi\gamma n_{0}}{k_{0}}J_{n}^{2}(R\sqrt{k_{0}^{2}-k_{z}^{2}}),\label{eq:app_decay}
\end{equation}
\begin{equation}
\Delta_{n}=\text{Im}(\Lambda_{n})=\frac{\pi\gamma n_{0}}{k_{0}}J_{n}(R\sqrt{k_{0}^{2}-k_{z}^{2}})Y_{n}(R\sqrt{k_{0}^{2}-k_{z}^{2}}).
\end{equation}
}Eq.(\ref{eq:app_decay}) shows that timed-Dicke state ($n=0$ and
$k_{z}=k_{0}$) $\beta(\varphi,z)=e^{ik_{0}z}$ has the fastest decay
rate $\text{Re}(\Lambda_{\text{TD}})=\pi\gamma n_{0}/k_{0}$. However,
collective Lamb shift for such state logarithmically diverges since
$Y_{0}(x)\approx(2/\pi)\ln(x/2)$ for small $x$. \textcolor{black}{For
the states with $R\sqrt{k_{0}^{2}-k_{z}^{2}}=A_{nl}$ where $A_{nl}$
is $l$th zero of the Bessel function $J_{n}(x)$, such as the state
$\beta_{n,k_{z}}(\varphi,z)=e^{in\varphi}e^{iz\sqrt{k_{0}^{2}-A_{nl}^{2}/R^{2}}}\approx e^{i(k_{0}-A_{nl}^{2}/2k_{0}R^{2})z}e^{in\varphi}$,
the decay rate and the collective Lamb shift vanish. }
\begin{figure}
\textcolor{black}{\includegraphics[width=1\columnwidth]{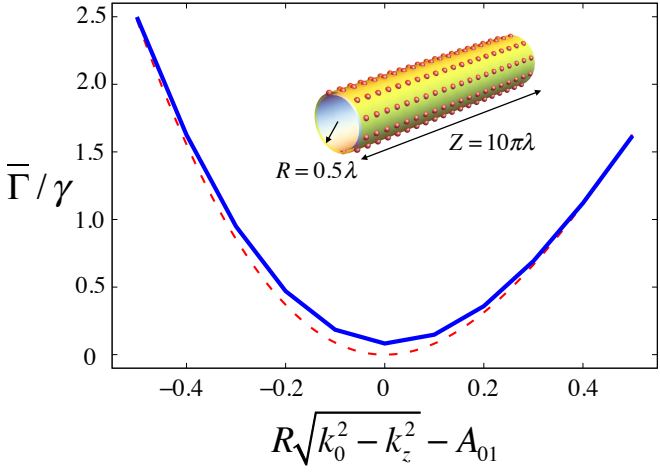}\caption{Solid line shows the average decay rate $\bar{\Gamma}$ of $\beta_{0,k_{z}}$
state for periodic distribution of atoms on cylindrical surface sketched
on the top. The cylinder consists of $1000$ atoms in $100$ layers
with $10$ atoms per each layer. Radius of the cylinder is $R=0.5\lambda$
and the distance between adjacent layers is $0.1\pi\lambda$. Analytical
result (\ref{eq:app_decay}) for an infinitely long cylindrical shell
with $100/\pi\lambda$ atoms per unit length is plotted as dashed
line. The horizontal axis is deviation of the $R\sqrt{k_{0}^{2}-k_{z}^{2}}$
from the root $A_{01}$ of the Bessel function $J_{0}(x)$. $\bar{\Gamma}=-\gamma\ln[P(1/\gamma)]$
is defined as average decay rate for time scale of $1/\gamma$.}
\label{fig:cylinder}}
\end{figure}
\textcolor{black}{In Fig.\ref{fig:cylinder}, we compare the decay
of axially symmetric atomic states for continuous and discrete distribution
of atoms on cylindrical surface. Namely, we plot the average decay
rate $\bar{\Gamma}=-\gamma\ln P(1/\gamma)$ of the state $\beta_{0,k_{z}}(\phi,z)=e^{ik_{z}z}$,
where $P(t)$ is the probability to find atoms excited, as a function
of $R\sqrt{k_{0}^{2}-k_{z}^{2}}-A_{01}$, where $A_{01}=2.404$ is
the first zero of $J_{0}(x)$. The average decay rate approaches zero
when $R\sqrt{k_{0}^{2}-k_{z}^{2}}=A_{01}$ for a discrete periodic
atomic distribution shown in Fig.\ref{fig:cylinder}. This agrees
with the analytical result in the continuous limit in Eq.(\ref{eq:app_decay})
plotted as a dashed line. Cylindrical atomic distribution can be achieved,
e.g., by adhering nano diamond with NV centers or SiV centers on a
carbon tube.}

\textcolor{black}{In summary, we demonstrate that the collective Lamb
shifts that are usually thought to destroy subradiance can be mitigated
by symmetry. For atomic distributions with mirror symmetry (a discrete
symmetry), the anti-symmetry states are subradiant, even when half
of the atoms are randomly distributed as long as the mirror symmetry
is maintained. Periodic distribution with intrinsic mirror symmetry
can be realized in ion traps and the subradiant anti-symmetry states
can be prepared by specially tailored anti-symmetric optical modes.
In addition, continuous symmetry can also be used to realize subradiance.}
\begin{figure}
\textcolor{black}{\includegraphics[width=1\columnwidth]{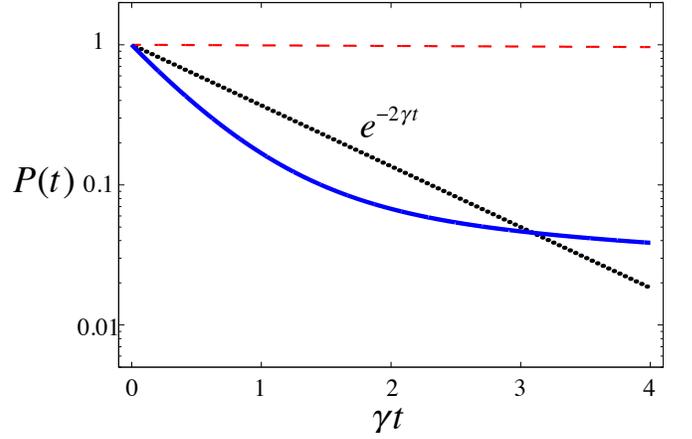}\caption{Results for vector field with $\Omega_{ij}^{v}$ and $\gamma_{ij}^{v}$.
The atomic distribution is the same as Fig.1 in main text. Probability
$P(t)$ to find atoms excited as a function of time for atoms initially
prepared in the $\ket-$ state. The solid curve takes the cooperative
Lamb shift into consideration. The dashed curve ignores the Lamb shift.
For comparison, we also plot single-atom decay curve $e^{-2\gamma t}$
(dot line). }
}

\textcolor{black}{\label{fig4}}
\end{figure}

\begin{acknowledgments}
\textcolor{black}{We gratefully acknowledge support of the National
Science Foundation Grant EEC-0540832 (MIRTHE ERC) and the Robert A.
Welch Foundation (Award A-1261). H. Cai is supported by the Herman
F. Heep and Minnie Belle Heep Texas A\&M University Endowed Fundheld/administered
by the Texas A\&M University.}
\end{acknowledgments}

\appendix

\section{\textcolor{black}{Results with vector field}}

\textcolor{black}{For the sake of simplicity, we use scalar field
theory throughout the main text, i.e.,
\begin{equation}
\Omega_{ij}=-\frac{\cos(k_{0}r_{ij})}{k_{0}r_{ij}}\gamma,
\end{equation}
\begin{equation}
\gamma_{ij}=\frac{\sin(k_{0}r_{ij})}{k_{0}r_{ij}}\gamma.
\end{equation}
However, if the polarization of electromagnetic field is considered,
$\Omega_{ij}$ and $\gamma_{ij}$ are\cite{GrossPhysRepo1982}
\begin{eqnarray}
\Omega_{ij}^{v} & = & \frac{3}{4}\gamma[-P_{ij}\frac{\cos kr_{ij}}{kr_{ij}}+Q_{ij}(\frac{\sin kr_{ij}}{(kr_{ij})^{2}}+\frac{\cos kr_{ij}}{(kr_{ij})^{3}})],\\
\Gamma_{ij}^{v} & = & \frac{3}{4}\gamma[P_{ij}\frac{\sin kr_{ij}}{kr_{ij}}+Q_{ij}(\frac{\cos kr_{ij}}{(kr_{ij})^{2}}-\frac{\sin kr_{ij}}{(kr_{ij})^{3}})],
\end{eqnarray}
where factors $P_{ij}=\hat{\mu}_{i}\cdot\hat{\mu}_{j}-(\hat{\mu}_{i}\cdot\hat{r}_{ij})(\hat{\mu}_{j}\cdot\hat{r}_{ij})$
and $Q_{ij}=\hat{\mu}_{i}\cdot\hat{\mu}_{j}-3(\hat{\mu}_{i}\cdot\hat{r}_{ij})(\hat{\mu}_{j}\cdot\hat{r}_{ij})$,
$\hat{r}_{ij}=\hat{r}_{i}-\hat{r}_{j}$. Here $\hat{r}_{j}$ and $\hat{\mu}_{j}$
are the position and dipole of the $j$th atom respectively. In Fig.4,
the features of the curve is the same as in Fig.1. Collective Lamb
shift $\Omega_{ij}^{v}$ significantly degrades the subradiance of
the state $\ket{-}$.}
\begin{figure}
\textcolor{black}{\includegraphics[width=1\columnwidth]{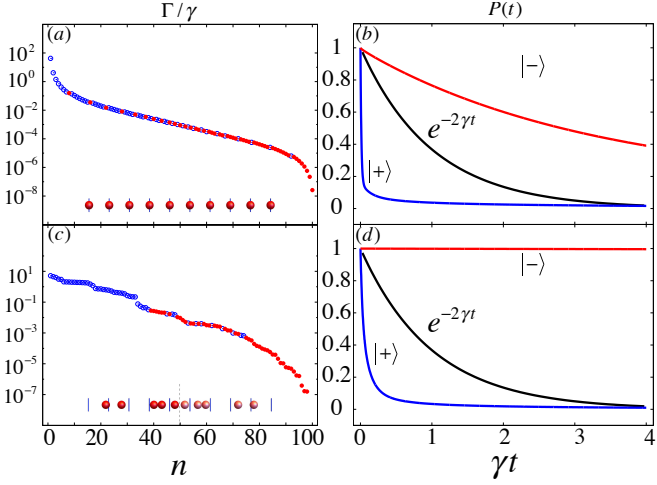}\caption{Eigenvalue distribution and population evolution for vector field.
(a) Distribution of decay rates for eigenstates of an ensemble of
$100$ atoms regularly placed along a line with spacing between adjacent
atoms $0.0006\lambda$. Blue empty dots are symmetric states while
red solid dots are anti-symmetric states. (b) Population decay of
states $\ket+$ and $\ket-$. Single-atom exponentially decaying curve
is shown for comparison. (c) and (d): The same as in (a) and (b),
but for random distribution of atoms with reflection symmetry.}
\label{fig2v}}
\end{figure}
\textcolor{black}{In Fig.\ref{fig2v}, we could restore subradiance
by taking advantage of the symmetry of atomic distribution, which
is similar to the result of Fig.2 with scalar field.}

\bibliographystyle{apsrev4-1}
\bibliography{/Users/HCai/Dropbox/my_cite_lib}

\end{document}